\documentclass[twocolumn,nofootinbib,aps,superscriptaddress,floatfix]{revtex4}

\usepackage{graphicx}

\newcommand{\beq}{\begin{equation}}
\newcommand{\eeq}{\end{equation}}
\newcommand{\beqa}{\begin{eqnarray}}
\newcommand{\eeqa}{\end{eqnarray}}
\newcommand{\nn}{\nonumber}

\newcommand{\Aslash}{{\cal A}\hspace*{-5.5pt}\slash}
\newcommand{\Bslash}{{\cal B}\hspace*{-5.5pt}\slash}
\def\Dslash{D\hspace*{-6.5pt}\slash}
\def\nslash{n\!\!\!\slash}
\def\bnslash{\bar n\!\!\!\slash}

\def\OMIT#1{{}}
\def\lqcd{\ensuremath{\Lambda_{\rm QCD}}}

\def\GeV{\mbox{GeV}}

\def\Bbar{\,\overline{\!B}{}}
\def\Kbar{\,\overline{\!K}{}}
\def\B0bar{\Bbar{}^0}
\def\K0bar{\Kbar{}^0}
\def\Ks0bar{\Kbar{}^{0*}}
\def\GeV{{\rm GeV}}
\def\d{{\rm d}}

\begin{document}

\preprint{\vbox{\hbox{UCSD/PTH 04--21} \hbox{LBNL--56648} \hbox{MIT--CTP 3571}
  \hbox{hep-ph/0412019}}}

\vspace*{1cm}

\title{\boldmath The photon polarization in $B\to X\gamma$ in the standard
model}

\author{Benjam\'\i{}n Grinstein}
\affiliation{Department of Physics, University of California at San Diego,
  La Jolla, CA 92093}

\author{Yuval Grossman}
\affiliation{Department of Physics, Technion--Israel Institute of Technology,
  Technion City, 32000 Haifa, Israel}

\affiliation{Physics Department, Boston University, Boston, MA 02215}

\affiliation{Jefferson Laboratory of Physics, Harvard University,
  Cambridge, MA 02138}

\author{Zoltan Ligeti}
\affiliation{Ernest Orlando Lawrence Berkeley National Laboratory,
  University of California, Berkeley, CA 94720}

\author{Dan Pirjol}
\affiliation{Center for Theoretical Physics, Massachusetts Institute for
  Technology, Cambridge, MA 02139}

\begin{abstract}

The standard model prediction for the $\B0bar\to X_{s,d}\gamma$ decay amplitude
with a right-handed photon is believed to be tiny, suppressed by $m_{s,d}/m_b$,
compared to the amplitude with a left-handed photon.  We show that this
suppression is fictitious: in inclusive decays, the ratio of these two
amplitudes is only suppressed by $g_s/(4\pi)$, and in exclusive decays by
$\lqcd/m_b$.  The suppression is not stronger in $\B0bar\to X_d \gamma$ decays
than it is in $\B0bar\to X_s \gamma$.  We estimate that the time dependent $CP$
asymmetries in $B\to K^*\gamma$, $\rho\gamma$, $K_S\pi^0\gamma$, and
$\pi^+\pi^-\gamma$ are of order 0.1 and that they have significant
uncertainties.

\end{abstract}

\maketitle

\section{Introduction}

The standard model (SM) predicts that photons are mainly left-handed in $b\to
q\gamma$ ($q = s, d$) decay (and right-handed in $\bar b\to \bar q\gamma$).  We
define the ratio
\beq \label{rdef}
r_{q}\, e^{i(\phi_q + \delta_q)} \equiv {A_R\over A_L}
  \equiv {A(\Bbar\to f_q\gamma_R)\over A(\Bbar\to f_q\gamma_L)} \,,
\eeq
where $\phi$ is a weak phase and $\delta$ is a strong phase.  It is usually
stated that $r_q = m_q/m_b \ll 1$ in the SM~\cite{Atwood:1997zr}. 
(Throughout this paper $\Bbar$ refers to $\B0bar$ or $B^-$ that contain a $b$
quark, and $r$, $\phi$ and $\delta$ depend on the final state, $f$.)  New
physics can modify this prediction, and therefore several methods have been
proposed to measure the photon helicity~\cite{Atwood:1997zr,Grossman:2000rk}.

In $B\to f\gamma$, where $f$ is a $CP$ eigenstate, since $\gamma_L$ and
$\gamma_R$ cannot interfere, the time dependent $CP$ asymmetry
\beqa
&&{} {\Gamma[\B0bar(t)\to f \gamma] - \Gamma[B^0(t)\to f\gamma]\over
   \Gamma[\B0bar(t)\to f\gamma] + \Gamma[B^0(t)\to f\gamma] } \nn\\
&&{}\qquad = S_{f\gamma} \sin(\Delta m\, t)
   - C_{f\gamma} \cos(\Delta m\, t)\,,
\eeqa
is sensitive to $r$.  In the SM, $\phi_s$ and $C_{f_s\gamma}$ are suppressed by
$|(V_{ub}V_{us})/(V_{tb}V_{ts})|$, and to first order in $r_s (\ll 1)$
\beq\label{Ss}
S_{f_s\gamma} = -2\, r_s \cos\delta_s \sin2\beta \,.
\eeq
The first measurements of such $CP$ asymmetries were carried out recently,
\beq
S_{K^*\gamma} = \cases{+0.25\pm0.63\pm0.14  & 
BABAR~\cite{Aubert:2004pe}\,,\cr
   -0.79^{+0.63}_{-0.50}\pm0.10  &  BELLE~\cite{Abe:2004xp}\,, }
\eeq
yielding a world average $S_{K^*\gamma} = -0.28 \pm 0.45$.  At a
super-$B$-factory the statistical error with 50\,ab$^{-1}$ data is estimated to
be $\delta(S_{K^*\gamma}) = 0.04$~\cite{Akeroyd:2004mj}.  The Belle
Collaboration also measured the $CP$ asymmetry $S_{K_S\pi^0\gamma} =
-0.58^{+0.46}_{-0.38} \pm 0.11$, integrating over the invariant mass range
$0.6\,\GeV < m_{K_S\pi^0} < 1.8\,\GeV$~\cite{Abe:2004sx}.  It will also be
possible to measure this $CP$ asymmetry in $B\to
\pi^+\pi^-\gamma$~\cite{Atwood:2004jj}, and maybe even with additional
pions~\cite{owen}.

The purpose of this paper is to study the SM prediction for $r$.  (For earlier
attempts to go beyond the naive estimate, see
Refs.~\cite{Grinstein:2000pc,Sehgal:2004xy}.)  We find that $r$ is only
suppressed by $g_s/(4\pi)$ in inclusive $b\to X\gamma$ decay, and by
$\lqcd/m_b$ in exclusive $\Bbar\to \Kbar^*\gamma$ and $\rho\gamma$ decay.

To understand the origin of such effects, recall that the effective
Hamiltonian for $b\to s\gamma$ is~\cite{GSW}
\beq\label{Heffs}
H_{\rm eff} = -{4G_F\over\sqrt2}\, V_{tb} V_{ts}^*\,
   \sum_{i=1}^8 C_i(\mu)\, O_i(\mu) \,.
\eeq
For our discussion the operators directly relevant are
\beqa\label{O27}
O_2 &=& (\bar c\, \gamma^\mu P_L b)\, (\bar s\, \gamma_\mu P_L c) \,,\nn\\
O_7 &=& {e\over16\pi^2}\, \bar s\, \sigma^{\mu\nu} F_{\mu\nu}
   (m_b P_R + m_s P_L)\, b \,,
\eeqa
where $P_{R,L} = (1\pm\gamma_5)/2$, and we neglect the $m_s P_L$ part of $O_7$
hereafter.  At the parton level, as long as $b\to s\gamma$ is a two-body decay
(either from the  leading contribution of $O_7$ or subleading virtual
contributions from $O_{i\neq7}$), the left-handed $s$ quark is back-to-back to
a photon.  Then the two-body kinematics implies that only $\gamma_L$ is
allowed. This argument does not apply to multi-body final states, such as $b
\to s \gamma\ +$ gluons.

The $m_b P_R$ part of the leading operator $O_7$ contributes only to $A_L$ to
all orders in the strong interaction.  To prove this, note that the
electromagnetic tensor for $\gamma_{L,R}$ is $F_{\mu\nu}^{L,R} = \frac12
(F_{\mu\nu} \pm i \widetilde F_{\mu\nu})$, where $\widetilde F_{\mu\nu} =
\frac12 \varepsilon_{\mu\nu\rho\lambda} F^{\rho\lambda}$, and $O_7$ can be
written in terms of $m_b F_{\mu\nu}^L$.  Thus, independent of hadronic physics,
the photon from $O_7$ is left-handed.  This argument only applies for $O_7$. 
Indeed, we find by explicit calculation that other operators produce
right-handed photons once QCD corrections are included.

\section{\boldmath Inclusive $B\to X_s\gamma$ and $B\to X_d\gamma$}

In this section we estimate $r$ from an inclusive calculation.  The result can
only be trusted if several hadronic final states are allowed to contribute, and
$r$ for specific final states cannot be obtained from this calculation model
independently.

\begin{figure}[t]
\centerline{\includegraphics[width=.3\columnwidth]{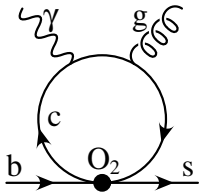}}
\caption{Leading contribution to $\Bbar\to X_s\gamma_R$.  A second diagram
with the gluon and the photon exchanged is not shown.}
\label{fig:O2brem}
\end{figure}

The leading contribution to the inclusive $\Bbar\to X_s\gamma_R$ rate is of
order $\alpha_s$.  It arises from bremsstrahlung contributions to the matrix
elements of operators other that $O_7$.  The numerically dominant contribution
comes from $O_2$ shown in Fig.~\ref{fig:O2brem}. The corresponding amplitude was
calculated in Ref.~\cite{Ali:1990tj,Pott:1995if}.  We find that it yields equal
rates for left- and right-handed photons at order $\alpha_s$, at any point in
the $b\to sg\gamma$ Dalitz plot. 

Because of the complicated $m_c$-dependence of the double differential rate,
$\d\Gamma_{22}^{\rm (brem)}/\d E_\gamma \d E_g$, we integrate over $E_g$ and
$E_\gamma$ numerically.  To reduce the large sensitivity to the scale of
$\alpha_s$, we include the known order $\alpha_s^2\beta_0$ contribution to
$\Gamma_{22}^{\rm (brem)}$~\cite{Ligeti:1999ea}.  (At this order the equality
of the decay rates to left- and right-handed photons is violated by less than
1\%, and can be neglected.)

Using the ``effective" Wilson coefficients at leading
order~\cite{Buras:1993xp}, $C_2(m_b) = 1.1$ and $C_7(m_b) = -0.31$,
$\alpha_s(m_b) = 0.22$ with $m_b=4.8\,$GeV, and $m_c=1.4\,$GeV, we obtain
\beq\label{22ratec}
{\Gamma_{22}^{\rm (brem)}\over \Gamma_0} \simeq 0.025 \,, \qquad
\Gamma_0 = {G_F^2 |V_{tb}V_{ts}^*|^2 \alpha_{\rm em}\, C_7^2\, m_b^5
  \over 32\pi^4}\,.
\eeq
This result corresponds to integrating the numerator over $x \equiv
2E_\gamma/m_b > 0.75$ (that is, roughly, $E_\gamma > 1.8\,$GeV).  This result
includes also the ${\cal O}(\alpha_s^2\beta_0)$ correction, which  is sizable,
indicating that the relevant scale of $\alpha_s$ may be well below $m_b$;
without including it the result in Eq.~(\ref{22ratec}) would be 0.015. Thus, we
find at lowest order in $g_s$
\beq\label{bigestimate}
\langle r_s \rangle \big|_{x > 0.75} 
  = \sqrt{\Gamma_{22}^{\rm (brem)} / (2 \Gamma_0)} \simeq 0.11\,.
\eeq
This value of $\langle r_s\rangle$ decreases only slowly with a stiffer cut on
$x$.

The value of $\cos\delta_s$ is physical, as it enters $S_{f_s\gamma}$ in
Eq.~(\ref{Ss}).  Yet it cannot be estimated from the inclusive calculation. 
The reason is that the dominant contribution to $A_R$ comes from the $b \to s
\gamma g$ amplitude generated by $O_2$, while to $A_L$ from the $b \to s
\gamma$ decay generated by $O_7$.  These are different final states, for which
one can chose the phase conventions independently.  These amplitudes can still
contribute to the same hadronic final states and interfere once hadronization
effects are included. Thus, the relevant phase for any final state is
determined by the hadronization processes and cannot be extracted from the
inclusive calculation.  Comparing the absorptive and dispersive parts of the
inclusive result, we find $\cos^2\delta_s \simeq 0.3$ with small variation over
$0.75 < x < 1$.  The only conclusion we can draw here is that we expect the
strong phase to be generically large.

Next we discuss inclusive $B\to X_d\gamma$. In Eq.~(\ref{Heffs}), operators
multiplying the suppressed CKM factor $V_{ub}V_{us}^*$ were neglected.  The
analogous terms are important in $b\to d\gamma$, since in this decay the $u$
quark loop is not CKM suppressed compared to the $c$ and $t$ loops.  The $b\to
d\gamma$ effective Hamiltonian is of the form
\beqa\label{Heffd}
H_{\rm eff} &=& -{4G_F\over\sqrt2}\, V_{tb} V_{td}^*
  \sum_{i=1}^8 C_i(\mu)\, O'_i(\mu) \\*
&& - {4G_F\over\sqrt2}\, V_{ub} V_{ud}^* 
  \sum_{i=1}^2 C_i(\mu)\, [O'_i(\mu) - O'_{i,u}(\mu)] \,, \nn
\eeqa
where prime denotes that in the standard operator basis in Eq.~(\ref{O27}) the
$s$ quark is replaced by $d$, and the $u$ subscript means that $c$ and $\bar c$
are replaced by $u$ and $\bar u$; for example,
\beq
O'_{2,u} = (\bar u\, \gamma^\mu P_L b)\, (\bar d\, \gamma_\mu P_L u) \,.
\eeq
It is useful to define
\beq
r_d\, e^{i(\phi_d + \delta_d)} = r_s\, e^{i \delta_s} 
  + {V_{ub} V_{ud}^* \over V_{tb} V_{td}^*}\, r_d^u\, e^{i \delta_d^u} ,
\eeq
where the first [second] term comes from the contribution of the top [bottom]
line in Eq.~(\ref{Heffd}) to $A_R$.  The first term is the same as in $b\to s$
decay (recall that we neglected~$m_s$).

For the $u$ quark loop,  taking the limit of the quark mass to zero, the
calculation of Ref.~\cite{Ligeti:1999ea} simplifies considerably, and we can
obtain analytic results
\beqa\label{22rateu}
{1\over\Gamma_0} {\d\Gamma_{22}^{(u, \rm brem)}\over \d x} 
&=& {C_2^2\over C_7^2}\, \bigg\{ {\alpha_s\over 27\pi}\, 2x(2-x) \\*
&+& {\alpha_s^2\beta_0\over 27 \pi^2}\, 
  \bigg[ {x(44 - 19x + 18x^2 - 24x^3)\over 12} \nn\\*
&&{} - {x(2 - x)\over 2} \ln[x(1 - x)] 
  + {2x^3\over 3} \ln x \bigg] \bigg\} , \nn
\eeqa
where $\alpha_s$ is evaluated at the scale $m_b$ in the $\overline{\rm MS}$
scheme.  For the difference of the rates to $\gamma_R$ and $\gamma_L$ we obtain
\beq\label{22rateudiff}
{1\over\Gamma_0} {\d\big[\Gamma_{22}^{(u,R)}-\Gamma_{22}^{(u,L)}\big]\over \d x}
  = {C_2^2\over C_7^2}\, {\alpha_s^2\beta_0\over 27 \pi^2}\, 
  {x^2(3 - 3x + 4x\ln x)\over 6} \,.
\eeq
This difference integrates to zero, and it gives a slight $\gamma_L$
enhancement near $x=1$.  We obtain for $x>0.75$
\beq\label{22rateunum}
{\Gamma_{22}^{\rm (u,brem)}\over \Gamma_0} \simeq 0.030\,.
\eeq

The absolute values of the amplitudes corresponding to the $c$ and $u$ loops
[the squares of which yield Eqs.~(\ref{22ratec}) and (\ref{22rateunum})] are
comparable to each other.  In the absence of strong phases this would result in
a cancellation and lead to a very small $r_d^u$.  Note that, as in the $b \to s
\gamma$ case, we cannot predict the sign of $\cos\delta_d^u$.  Moreover, the
values of $r_d^u$ and $\delta_d^u$ are sensitive to the difference between the
strong phases of the $c$ and $u$ loops, for which we do not consider the
perturbative result reliable.  In particular, the matrix element of $O'_{2,u}$
in $b\to d\gamma$ may have sizable long distance contributions.  In any event,
the short distance calculation predicts that the strong phase vanishes for the
$u$ loop contribution, while it is sizable for the $c$ loop.  Therefore, the
cancellation in $r_d^u$ is unlikely to be effective.  The important point is
that we expect $r_d \sim r_d^u \sim r_s$.

The crucial difference between the time dependent $CP$ asymmetries in $B\to
f_s\gamma$ and $B\to f_d\gamma$ is that in the latter case, naively, there are
two very strong suppression factors.  First, considering $O_7$ only,
$S_{f_d\gamma}$ is suppressed by $m_d/m_b$.  Second, the phase of the dominant
decay amplitude, $V_{tb} V_{td}^*$, cancels the phase of the $B^0-\B0bar$
mixing amplitude, yielding another strong suppression of $S_{f_d\gamma}$.  Both
of these suppressions are fictitious, since the $V_{ub} V_{ud}^*\, (O'_2 -
O'_{2,u})$ contributions lift both the $m_d/m_b$ suppression, just like in the
$b\to s\gamma$ case discussed earlier, and also the suppression coming from the
cancellation of the mixing and the decay phases.  The leading contribution to
$S_{f_d\gamma}$ is proportional to $r_d^u$, which gives a contribution to $A_R$
with weak phase $\gamma$. Using the fact that the phase of the mixing amplitude
is $2 \beta$, and that of $A_L$ is $\beta$, we obtain at leading order
\beq\label{Sd}
S_{f_d\gamma} = -2\, \bigg| {V_{ub} V_{ud}^* \over V_{tb} V_{td}^*}\bigg|\, 
  r_d^u \cos\delta_d^u \sin(\beta+\gamma) \,.
\eeq
This result is independent at first order in $r_d^u$ of the small direct $CP$
violation in $b\to d\gamma$ in the SM.

A model dependent way to connect the inclusive calculations with exclusive
$B\to K^*\gamma$ or $\rho\gamma$ is the Ali--Greub model~\cite{Ali:1990vp}
obtained by smearing the inclusive rate with a model shape function and
integrating over $E_\gamma > (m_B^2-1\,\GeV^2)/(2m_B)$ (i.e., attributing the
rate to $m_X < 1\,$GeV to the $K^*$ or $\rho$).  Restricting the shape function
parameters to reproduce ${\cal B}(B\to K^*\gamma)$, we obtain $r_{K^*} \sim
0.025$, with little sensitivity to the model parameters.  This is comparable in
magnitude but independent of the $m_s/m_b$ contribution.  This model also
yields $r_\rho \sim r_{K^*}$, which is much larger than $m_d/m_b$.  To compare
with the ``more inclusive" measurement~\cite{Abe:2004sx}, we computed $\langle
r_s \rangle_{m_{X_s} < 1.8\,\GeV} \sim 0.055$ using simple shape function
models.  Given the model uncertainties, this figure should be taken as a very
rough estimate.

\section{\boldmath Exclusive $B\to K^*\gamma$ and $B\to \rho\gamma$}

We consider next the photon polarization in exclusive $\Bbar\to \Kbar^*\gamma$
and $\rho\gamma$ decays using SCET~\cite{Bauer:2000ew}.  We prove that
$O_{1-6}$ contribute only to $A_L$ to all orders in $\alpha_s$ at leading order
in $\lqcd/m_b$ (this is viewed as of the same order as $m_{K^*}/E_{K^*}$).  We
identify several types of subleading SCET operators that give amplitudes to
right-handed photons, and yield $r_{K^*}$ and $r_\rho$ of order $\lqcd/m_b$.

Using SCET, a factorization theorem for heavy-to-light form factors has been
proven at leading order in
$\lqcd/m_b$~\cite{Bauer:2002aj,Beneke:2003pa,Lange:2003pk}.  There are two
contributions to the form factors, a nonfactorizable (or soft or form-factor)
part and factorizable (or hard scattering) part, which are of the same order in
$\lqcd/m_b$.

The exclusive radiative decays considered here were analyzed in an expansion in
$\lqcd/m_b$ in~\cite{Grinstein:2000pc,Beneke:2001at,Bosch:2001gv,Ali:2001ez},
and recently in SCET~\cite{Chay:2003kb}.  The nonfactorizable part receives its
dominant contribution from $O_7$ and gives rise to $A_L$ only.  The operator
$O_2$ enters at leading order only via factorizable contributions.

When operators $O_{i\neq7}$ are included, the hierarchy of the relevant scales
gets rather complicated, since $m_b^2 > m_c^2 \sim \lqcd m_b > \lqcd^2$.  
Ref.~\cite{Chay:2003kb} assumed that the $c\bar c$ loop is dominated by hard
loop momenta and can be integrated out near the scale $m_b$.  Here we adopt a
simpler approach, by neglecting the charm mass and assuming that its effects
can be included as a perturbation using the formalism of
Ref.~\cite{Leibovich:2003jd} without encountering singularities.  This
assumption is borne out by the explicit one- and two-loop calculations.

The effective Hamiltonian in Eq.~(\ref{Heffs}) is matched in SCET$_{\rm I}$
onto
\beqa\label{HeffSCET}
H_{\rm eff} &=& \frac{G_F V_{tb} V_{ts}^*\, e}{\sqrt2\, \pi^2}\,
  E_\gamma \Big[ c(\omega)\, \bar s_{n,\omega} \Aslash^\perp m_b P_L\, b_v \\
&+& b_{1L}(\omega_i)\, O^{(1L)}(\omega_i)
  + b_{1R}(\omega_i)\, O^{(1R)}(\omega_i) + {\cal O}(\lambda^2) \Big] , \nn
\eeqa
where the $\omega$'s are the usual collinear label momenta.  The relevant modes
are soft quarks and gluons with momenta $k_s \sim \Lambda$ and two types of
collinear quarks and gluons along $n$ and $\bar n$ (including charm, which can
be soft or collinear).  Note that the approach of Ref.~\cite{Chay:2003kb},
treating $m_c \sim m_b$, will likely require additional modes.  We take the
photon momentum as $q_\mu = E_\gamma \bar n_\mu$, the collinear $s$ quark to
move along $n_\mu$, and ${\cal A}_\mu^\perp$ denotes the transverse photon
field.  The operator in the first line in Eq.~(\ref{HeffSCET}) occurs at
leading order in the expansion parameter, $\lambda = \sqrt{\Lambda/m_b}$, and
its Wilson coefficient is dominated by $C_7$, $c(\omega) = C_7 + {\cal
O}[\alpha_s(m_b)]$.  The operators in the second line,
\beqa
O^{(1L)}(\omega_1,\omega_2) &=& \bar s_{n,\omega_1}\, \Aslash^\perp 
  \Big[\frac1{\bar n\cdot {\cal P}} ig \Bslash^\perp_n\Big]_{\omega_2}
  P_R\, b_v , \nn\\*
O^{(1R)}(\omega_1,\omega_2) &=& \bar s_{n,\omega_1}
  \Big[\frac1{\bar n\cdot {\cal P}} ig \Bslash^\perp_n\Big]_{\omega_2}
  \Aslash^\perp  P_R\, b_v ,
\eeqa
are the only SCET$_{\rm I}$ operators suppressed by $\lambda$ that couple to a
transverse photon and are allowed by power counting and $s$ chirality.  Here
$ig{\cal B}_n^\nu \equiv [\bar n\cdot iD_c\,, iD_{c\perp}^\nu]$ is the
collinear gluon field tensor; for the remaining notations see Ref.~\cite{ps1}. 
The operators $O^{(1L)}$ and $O^{(1R)}$ couple only to $\gamma_{L,R}$,
respectively. Their Wilson coefficients are
\beqa\label{b1LR}
b_{1L}(\omega_1,\omega_2) &=& C_7 + C_2/3 + {\cal O}[C_{3-6}, \alpha_s(m_b)]\,,
  \nn\\*
b_{1R}(\omega_1,\omega_2) &=& -\, C_2/3 + {\cal O}[C_{3-6}, \alpha_s(m_b)]\,.
\eeqa

Although the operators in the first and second lines of Eq.~(\ref{HeffSCET})
are of different orders in $\lambda$, after matching onto SCET$_{\rm II}$, they
contribute at the same order in $\lqcd/m_b$ to the $B\to K^*\gamma$ amplitude. 
The leading order $c(\omega)$ term gives the nonfactorizable contribution,
which only contributes to $A_L$, while $O^{(1L)}$ and $O^{(1R)}$ give the
factorizable contributions through time-ordered products with the
ultrasoft-collinear Lagrangian, ${\cal L}_{\xi q}^{(1)}$~\cite{Beneke:2002ph}. 
After matching onto SCET$_{\rm II}$ one finds, schematically
\beqa
&& \int\! \d^4 x\, T \big\{ O^{(1L)}(\omega_i), i{\cal L}_{\xi q}^{(1)}(x)\big\}
  \to \int\! \d k_+\, {\cal J}_\perp(\omega^{(\prime)}_i, k_+) \nn\\
\label{O1L}
&&{} \quad \otimes \big[(\bar qY_n)_{k_+} \nslash \Aslash^\perp 
  \gamma^\alpha_\perp P_R (Y^\dagger_n b_v)\big]\,
  (\bar s_{n,\omega'_1} \bnslash\gamma_\perp^\alpha q_{n,\omega'_2}) ,\\
&& \int\! \d^4 x\, T \big\{ O^{(1R)}(\omega_i), i{\cal L}_{\xi q}^{(1)}(x)\big\}
  \to \int\! \d k_+\, {\cal J}_\parallel(\omega^{(\prime)}_i, k_+) \nn\\
\label{O1R}
&&{} \quad \otimes \big[(\bar qY_n)_{k_+} \nslash \Aslash^\perp P_R 
  (Y^\dagger_n b_v)\big]\, (\bar s_{n,\omega'_1} \bnslash P_L q_{n,\omega'_2}) ,
\eeqa
where ${\cal J}_{\perp,\parallel}$ are jet functions that have expansions in
$\alpha_s(\sqrt{\lqcd m_b})$.  The operator in Eq.~(\ref{O1L}) contributes to
$B\to K^* \gamma$ at leading order,
\beqa
&&\!\!\!\! A_L = \frac{G_F V_{tb} V_{ts}^*\, e}{\sqrt2\, \pi^2}\, {m_B^3\over 2}
  \Big[ c(m_B) \zeta_\perp^{BK^*}
  + {f_B f_{K^*}^\perp\over m_B}\! \int\! \d x \d t \d k_+  \nn\\
&&\!\!\!\! \times \, b_{1L}(m_B (1-t), m_B t)\,
  {\cal J}_\perp(x,t,k_+) \phi_{K^*}^\perp(x)\, \phi_B^+(k_+)\Big] ,
\eeqa
where all nonperturbative matrix elements are defined as in~\cite{ps1}.  The
operator in Eq.~(\ref{O1R}) only gives rise to longitudinal $K^*$ and therefore
does not contribute to $B\to K^*\gamma$.  This proves that $O_2$ contributes at
leading order in $\lqcd/m_b$ to $\Bbar\to \Kbar^*\gamma_L$ via Eq.~(\ref{O1L}),
but its contribution to $\Bbar\to \Kbar^*\gamma_R$ vanishes at this order. 
(Interestingly, it would contribute to $\Bbar^*\to \Kbar^{(*)}\gamma_R$ at
leading order.)  The same proof also holds for the other four-quark operators,
$O_{1-6}$.

This result agrees with the ${\cal O}(\alpha_s)$
computation~\cite{Beneke:2001at,Bosch:2001gv}, and extends it to all orders in
$\alpha_s$.  This is important, since $\alpha_s(\mu)$ at the hard-collinear
scale $\mu^2 \sim \Lambda m_b$ may or may not be
perturbative~\cite{Bauer:2002aj,Bauer:2004tj,Hill:2004if}.  The suppression of
$A_R$ can be understood from a simple helicity argument.  Since the $s$ quark
is left-handed, the $K^*$ cannot be right-handed, unless additional
right-handed gluons end up in the final meson.  However, the contributions of
higher Fock states are power suppressed.

The $\Bbar\to \Kbar^*\gamma_R$ amplitude does arise at subleading order in
$\lqcd/m_b$.  There are several sources of such corrections.  For example, (i)
time ordered products of $O^{(1R)}$ with the subleading collinear Lagrangian
${\cal L}_{\xi\xi}^{(n\geq 1)}$, which lead to factorizable contributions
similar to Eq.~(\ref{O1L}) containing an explicit factor
$\alpha_s(\sqrt{\lqcd m_b})$; (ii) higher order terms in the SCET$_{\rm I}$
effective Lagrangian in Eq.~(\ref{HeffSCET}), such as the ${\cal O}(\lambda^2)$
operator
\beqa\label{O2R}
O^{(2R)}(\omega) &=& \int \frac{\d x_+}{2\pi}\, \d k_-\, e^{-i x_+ k_-/2}\, 
  \kappa\Big({2E_\gamma k_-\over m_c^2}\Big) \nn\\
&\times& \bar s_{n,\omega} 
  \big[Y^\dagger_{\bar n} i\Dslash^{\rm\,us} Y_{\bar n}\big]\!
  \Big({\bar n x_+\over2}\Big)\, \Aslash_\perp P_R\, b_v ,
\eeqa
where the gauge invariant operator $[Y^\dagger_{\bar n} i\Dslash^{\rm\,us}
Y_{\bar n}](x)$ contains Wilson lines of the ultrasoft fields in the $\bar n$
direction.  $O^{(2R)}$ is obtained by matching the graph in Fig.~1 with one
ultrasoft gluon, and its Wilson coefficient is related to $b_{1R}$ in
Eq.~(\ref{b1LR}) by reparameterization invariance~\cite{Manohar:2002fd}.  The
$C_2/3$ terms in Eq.~(\ref{b1LR}) correspond to $m_c = 0$, when $\kappa(z) =
1/2$.  In this case the $k_-$ and $x_+$ integrals in Eq.~(\ref{O2R}) can be
performed trivially.  However, in Eq.~(\ref{O2R}) we kept $m_c \neq 0$ to
exhibit the analogy with the nonperturbative corrections to inclusive $b\to
s\gamma$ decay~\cite{Voloshin:1996gw,Ligeti:1997tc}.  For $m_c \neq 0$,
additional nonlocality is introduced by $\kappa(z) = 1/2 - 2
\arctan[\sqrt{z/(4-z)}]^2 / z$ with $z = 2E_\gamma
k_-/m_c^2$~\cite{Ali:1990tj}.  [For $m_c\neq0$, the $C_2/3$ terms in
Eq.~(\ref{b1LR}) should be multiplied by $2\kappa(2E_\gamma\omega_2/m_c^2)$.] 
Expanding $O^{(2R)}$ in $x_+$ reproduces the series of operators proportional to
$(m_b\lqcd/m_c^2)^n$ in Eq.~(5) in \cite{Ligeti:1997tc}.  Time-ordered products
of $O^{(2R)}$ with the ultrasoft-collinear Lagrangian, ${\cal L}_{\xi q}^{(2)}$,
give ${\cal O}(\lqcd/m_b)$ soft contributions to $A_R$, with no $\alpha_s$
suppression compared to $A_L$.

A complete study of these subleading contributions is rather involved and we
leave it for future work.  It seems unlikely to us that a cancellation could
result in a suppression of $r_{K^*}$ and $r_\rho$ to order $\lqcd^2/m_b^2$. 
This leads to the dimensional estimate
\beq\label{excls}
r_{K^*} \sim \frac13\, \frac{C_2}{C_7}\, \frac{\lqcd}{m_b} \sim 0.1\,.
\eeq
This effect dominates over the $m_s$ piece of $O_7$ for $r_{K^*}$.  The
estimate for $r_\rho$ is more involved because of the contributions with
different weak phases,
\beq\label{excld}
r_\rho \sim r_{K^*} \bigg[ 1 + {V_{ub} V_{ud}^* \over V_{tb} V_{td}^*} 
  \bigg( C_{\rm loop}\, {m_c^2\over m_b^2} +
  C_{\rm WA}\, 4\pi\, {\lqcd \over m_b} \bigg) \bigg] ,
\eeq
The $C_{\rm loop}$ term comes from the non-cancellation of the $c$ and $u$
loops, and we expect it to have a numerically large coefficient.  The $C_{\rm
WA}$ term arises from weak annihilation, whose contribution to $A_R$ at order
$\lqcd/m_b$ vanishes~\cite{Grinstein:2000pc}. The latter contributes
significantly only in $B^\pm$, while in $B^0$ it is color suppressed.  Thus, we
expect that the SM prediction for $S_{\rho\gamma}$ is not much smaller than it
is for $S_{K^*\gamma}$.

For higher mass one-body hadronic final states, $A_R$ still vanishes at leading
order, but the suppression by $m_X/E_X$ is expected to be less effective as
$m_X$ increases (although there is no evidence for this in the $B\to DX$
data~\cite{Ligeti:2001dk}).  Thus, the SM value of $r$ is expected to depend on
the final state.  For high-mass and multi-body final states $A_R$ may arise,
formally, at leading order in $\lqcd/m_b$.  For example, $\Bbar\to \Bbar^*
\pi_{\rm (soft)}$ followed by $\Bbar^*\to \Kbar^{(*)}\gamma_R$ can give rise to
$\B0bar\to K_S\pi^0\gamma_R$ with $m_{K_S\pi^0} \sim \sqrt{\lqcd m_b}$, without
a $\lqcd/m_b$ suppression.  Therefore, averaging the results of $B\to
K^*\gamma$~\cite{Aubert:2004pe,Abe:2004xp} with $B\to K_S\pi^0
\gamma$~\cite{Abe:2004sx} is not free from theoretical uncertainties.

\section{Discussion and Summary}

We studied the standard model prediction for the $\Bbar\to X_{s,d}\gamma_R$
decay amplitude with a right-handed photon, $A_R$, compared to the amplitude
with a left-handed photon.  Considering only $O_7$, their ratio is $r_{s,d} =
m_{s,d}/m_b$, which reproduces the often-quoted prediction independent of the
hadronic final state.  However, including the other operators, $O_{i \neq 7}$,
$A_R$ becomes much larger than this naive estimate, and hadronic physics gives
rise to sizable uncertainties.  The time dependent $CP$ asymmetries also become
sensitive to the strong phase.

In inclusive $B\to X_s\gamma$ and $X_d\gamma$ decays, $A_R$ is only suppressed
by $g_s/(4\pi)$.  We calculated $r$ inclusively, and found it to be of order
$0.1$ depending on the cut on the photon energy [see Eq.~(\ref{bigestimate})]. 
While this calculation is reliable, it cannot be easily compared to data.  If
one restricts the hadronic final state, such as in the measurement of the time
dependent $CP$ asymmetry $S_{K_S\pi^0\gamma}$ with a range of $K_S\pi^0$
invariant masses, then that can no longer be calculated using inclusive
methods.  Still, our results indicate that it would be hard to argue that a
measurement of $S_{K_S\pi^0\gamma} \sim 0.1$ cannot be due to SM physics.

In exclusive $\Bbar\to \Kbar^*\gamma$ and $\rho\gamma$ decays, we proved using
SCET that $A_R$ vanishes at lowest order in the $\lqcd/m_b$ expansion to all
orders in $\alpha_s$.  The leading contribution to $A_R$ is formally of order
$\lqcd/m_b$, but numerically it is enhanced by $C_2/C_7$ [see
Eqs.~(\ref{excls}) and (\ref{excld})].  The result depends on unknown hadronic
matrix elements, which give rise to a sizable uncertainty in the SM prediction.

Both the inclusive and the exclusive calculations predict that $r_s$ and $r_d$
are comparable to each other in the SM.  Thus, we also expect the time
dependent $CP$ asymmetries in $\B0bar\to f_d \gamma$ to be comparable to those
in $\B0bar\to f_s \gamma$ (except for a modest suppression by
$|V_{ub}/V_{td}|$), in contrast to what has been widely believed.  We estimate
these asymmetries in decays such as $B\to K^*\gamma$, $\rho\gamma$,
$K_S\pi^0\gamma$, and $\pi^+\pi^-\gamma$ to be of order $0.1$, with large
uncertainties.

To conclude, our main result is that the standard model prediction for
$A_R/A_L$ is of order $0.1$, with sizable uncertainty.  A measurement of it
much above this level would indicate new physics.  To draw conclusions from a
smaller value would require a more complete analysis and knowledge of hadronic
matrix elements including strong phases.  More effort in this direction would
be welcome.

\acknowledgments

We thank Andy Cohen and Iain Stewart for helpful discussions.
Z.L.\ thanks the particle theory group at Boston University for its hospitality
while part of this work was completed.
B.G.~was supported in part by the DOE under Grant DE-FG03-97ER40546.
Z.L.~was supported in part by the Director, Office of Science, Office of
High Energy and Nuclear Physics, Division of High Energy Physics, of the U.S.\
Department of Energy under Contract DE-AC03-76SF00098 and by a DOE Outstanding
Junior Investigator award.
D.P.\ was supported by the U.S.\ Department of Energy under Grant
DOE-FG03-97ER40546 and by the NSF under grant PHY-9970781.

\end{document}